\documentclass[twocolumn,usenames,dvipsnames,twocolappendix]{aastex63}

\usepackage{amsmath}

\shorttitle{From UV-bright Galaxies to Early Disks}
\shortauthors{Semenov, Conroy, Hernquist}
\graphicspath{{./}{figures/}}

\usepackage{etoolbox}
\makeatletter
\patchcmd{\NAT@citex}
  {\@citea\NAT@hyper@{\NAT@nmfmt{\NAT@nm}\NAT@date}}
  {\@citea\NAT@nmfmt{\NAT@nm}\NAT@hyper@{\NAT@date}}
  {}% Do nothing if patch works
  {}% Do nothing if patch fails
\patchcmd{\NAT@citex}
  {\@citea\NAT@hyper@{%
     \NAT@nmfmt{\NAT@nm}%
     \hyper@natlinkbreak{\NAT@aysep\NAT@spacechar}{\@citeb\@extra@b@citeb}%
     \NAT@date}}
  {\@citea\NAT@nmfmt{\NAT@nm}%
   \NAT@aysep\NAT@spacechar%
   \NAT@hyper@{\NAT@date}}
  {}% Do nothing if patch works
  {}% Do nothing if patch fails
\patchcmd{\NAT@citex}
  {\@citea\NAT@hyper@{%
     \NAT@nmfmt{\NAT@nm}%
     \hyper@natlinkbreak{\NAT@spacechar\NAT@@open\if*#1*\else#1\NAT@spacechar\fi}%
       {\@citeb\@extra@b@citeb}%
     \NAT@date}}
  {\@citea\NAT@nmfmt{\NAT@nm}%
   \NAT@spacechar\NAT@@open\if*#1*\else#1\NAT@spacechar\fi%
   \NAT@hyper@{\NAT@date}}
  {}% Do nothing if patch works
  {}% Do nothing if patch fails
\makeatother

\def\eth{e_{\rm th}}

\def\cs{c_{\rm s}}

\def\vcirc{v_{\rm circ}}

\def\Mgas{M_{\rm gas}}
\def\Msf{M_{\rm sf}}

\def\Rcirc{R_{\rm circ}}

\def\jzjc{j_z/j_{\rm c}}

\def\rhoSFR{\dot{\rho}_{\star}}
\def\SFR{\dot{M}_{\star}}

\def\tff{t_{\rm ff}}
\def\tauff{\tau_{\rm ff}}
\def\taudep{\tau_{\rm dep}}

\def\epsff{\epsilon_{\rm ff}}
\def\avir{\alpha_{\rm vir}}
\def\stot{\sigma_{\rm tot}}
\def\st{\sigma_{\rm turb}}
\def\eturb{e_{\rm turb}}
\def\nsf{n_{\rm sf}}

\def\pc{{\rm \;pc}}

\def\kms{{\rm \;km\;s^{-1}}}

\def\Myr{{\rm \;Myr}}
\def\Gyr{{\rm \;Gyr}}
\def\cc{{\rm \;cm^{-3}}}
\def\K{{\rm \;K}}

\defcitealias{semenov24a}{Paper~I}

\newcommand{\newtext}{}

\begin{document}

\title{From UV-bright Galaxies to Early Disks: \\the Importance of Turbulent Star Formation in the Early Universe}

\author[0000-0002-6648-7136]{Vadim A. Semenov}
\altaffiliation{\href{mailto:vadim.semenov@cfa.harvard.edu}{vadim.semenov@cfa.harvard.edu}}
\affiliation{Center for Astrophysics $|$ Harvard \& Smithsonian, 60 Garden St, Cambridge, MA 02138, USA}

\author[0000-0002-1590-8551]{Charlie Conroy}
\affiliation{Center for Astrophysics $|$ Harvard \& Smithsonian, 60 Garden St, Cambridge, MA 02138, USA}

\author[0000-0001-6950-1629]{Lars Hernquist}
\affiliation{Center for Astrophysics $|$ Harvard \& Smithsonian, 60 Garden St, Cambridge, MA 02138, USA}

\begin{abstract}
Bursty star formation at early times can explain the surprising abundance of early UV-bright galaxies revealed by JWST but can also be a reason for the delayed formation of galactic disks in high-resolution cosmological simulations. We investigate this interplay in a cosmological simulation of an early-forming Milky Way analog with detailed modeling of the cold turbulent interstellar medium (ISM), star formation, and feedback. We find that the modeling of locally variable star formation efficiency (SFE) coupled with the ISM turbulence on the scales of star-forming regions is important for producing both early bursty evolution and early formation and survival of galactic disks. 
Such a model introduces a qualitatively new channel of the global star formation rate (SFR) burstiness caused by chaotic fluctuations in the average SFE due to changes in the ISM turbulence, which, in our simulation, dominates the short-term SFR variability.
The average SFE stays low, close to $\sim 1\%$ per freefall time, and its variation decreases when the gas disk forms, leading to only mild effects of stellar feedback on the early disk, enabling its survival. By rerunning our simulation with fixed SFE values, we explicitly show that low SFEs lead to smoother SFR histories and disk survival, while high SFEs lead to bursty SFRs and hinder disk formation. The model with variable SFE switches between these two regimes at the moment of disk formation. These trends are missing in more commonly used star formation prescriptions with fixed SFE; in particular, the prescriptions tying star formation to molecular gas should be interpreted with caution because the two are decoupled at early times, as we also show in this paper.
\end{abstract}

%-----------------------------------------------------------------
%-----------------------------------------------------------------

%-----------------------------------------------------------------
%-----------------------------------------------------------------
\section{Introduction}
%-----------------------------------------------------------------

With the advent of ALMA and JWST our paradigm of galaxy formation has shifted. Early JWST results have revealed a surprisingly high abundance of UV-bright galaxies in the very early Universe \citep[e.g.,][]{finkelstein22,naidu22,boylan-kolchin23,labbe23}. Many of these early galaxies exhibit disk-like morphologies \citep[e.g.,][]{ferreira22a,ferreira22b,jacobs22,naidu22,nelson22,robertson23,kuhn24}. 
These findings corroborated the idea that disk galaxies likely form earlier than previously thought, at $z > 4$, as also suggested by mounting observations of dynamically cold gaseous disks at these high redshifts by ALMA \citep[e.g.,][]{smit18,neeleman20,pensabene20,rizzo20,rizzo21,fraternali21,lelli21,tsukui-iguchi21,herreracamus22,posses23,romanoliveira23}. Also in line with this picture, recent discoveries from the local Galactic archeological data facilitated by spectroscopic and astrometric surveys of nearby stars (such as APOGEE, LAMOST, Gaia, H3) suggest that the disk of our own Milky Way (MW) formed surprisingly early, within the first $\sim$2 Gyr of evolution \citep{bk22,bk24,conroy22,rix22,xiang-rix22,nepal24}.

Such early buildup of galaxies and the formation and survival of disks pose a challenge for the theory of galaxy formation. On the one hand, detailed investigations of the abundances of early galaxies in cosmological simulations and analytic models that followed the JWST discoveries showed that the high end of the rest-frame-UV luminosity functions can be explained by the stochastic variability of star formation rates (SFRs) at early times \citep[e.g.,][]{keller23,mason23,mccaffrey23,sun23-fire,shen23,kravtsov-belokurov24}. In this scenario, the UV luminosities of galaxies that experience a recent burst of star formation are upscattered, resulting in a more populated high-luminosity tail, alleviating the tension with observational data. To explain the observed luminosity functions, the SFR variability must increase at $z > 10$ and settle down at later times, suggesting a transition between different regimes of galaxy formation \citep[e.g.,][]{shen23,kravtsov-belokurov24}.

On the other hand, cosmological simulations exhibiting strong early variation of SFRs appear to be in tension with the picture of early disk formation \citep[e.g.,][]{bk22,mccluskey23}. Strong variations of the SFR and associated feedback energy and momentum injection can perturb or even destroy the early disk in the process of formation \citep[e.g.,][]{agertz16}, leading to a delayed formation of galactic disks. 

Whether the star formation and feedback in the real early Universe are as bursty as suggested by the above results remains a subject of active debate. Alternative explanations for the high abundance of early UV-bright galaxies have been proposed in the recent literature, including variations in the IMF shape \citep[e.g.,][]{shen23,yung24} and dust abundances \citep[e.g.,][]{ferrara23} and stellar feedback being inefficient at early times \citep[e.g.,][]{dekel23,qin23,boylan-kolchin24}. 

Similarly, the early formation and survival of the MW disk can be explained by its unusual assembly history. Based on the TNG50 cosmological-volume simulation, only $\sim$10\% of MW-mass disk galaxies form their disks as early as the local Galactic archeology data suggests \citep[][see also \citealt{dillamore24}]{semenov23a,semenov23b}. However, given that the effective equation of state (EoS) approach used in TNG50 leads to inherently smoothed star formation histories, the interplay between the mass assembly history and star formation burstiness in disk formation and evolution remains an open question.

Overall, the above trade-off between the early burstiness of SFR favored by the observed high-redshift luminosity functions and the early formation of galactic disks makes it important to investigate both simultaneously. In \citet[][hereafter, \citetalias{semenov24a}]{semenov24a}, we show that switching from the effective EoS prescription in TNG to explicit modeling of the cold turbulent interstellar medium (ISM) leads to both significantly earlier disk formation and stronger variation of the SFR at early times. This implies that the strong SFR burstiness and early disk spinup are not mutually exclusive. In this paper, we investigate this effect in more detail and highlight the importance of explicit modeling of turbulence-regulated local star formation efficiency (SFE) per freefall time.

Over the past decade, an increasing number of galaxy formation simulations have been using such star formation models with continuously variable SFE \citep[e.g.,][]{braun15,semenov16,semenov21-tf,trebitsch17,rosdahl18,kretschmer20,gensior20,vintergatan1,vintergatan4}.  
This approach is qualitatively different from more common star formation prescriptions that treat SFE as a tunable parameter, introduce star formation thresholds, and, sometimes, modulate the SFR density with the local abundance of molecular gas. Such prescriptions are motivated by the star formation scaling relations observed in nearby star-forming galaxies, which suggest near-universal values of SFE per freefall time of the order of 1\% \citep[e.g.,][]{evans09,evans14,heiderman10,lada10,lee16,leroy17,utomo18} and a near-linear correlation between SFR and molecular gas surface density on $\gtrsim$kiloparsec scales \citep[molecular Kennicutt--Schmidt relation; e.g.,][]{wong02,bigiel08,bigiel11,leroy13,bolatto17}. 
However, it is not clear whether such relations hold in the extreme regimes of the early Universe. The explicit forward modeling of local SFE and, independently, molecular gas in our simulations enables us to make predictions about the local SFE values and the relation between molecular and star-forming gas reservoirs at early times, which we also investigate in this paper.

The ISM, star formation, and feedback model that we use in this paper is particularly appealing for this purpose because it can produce a realistic ISM structure in a range of environments in which the model has been tested in idealized simulations of isolated galaxies. In particular, the model produces a realistic correlation between molecular gas and SFR on kiloparsec and larger scales \citep{semenov17,semenov18,semenov19} and the spatial correlation between the two on subkiloparsec scales \citep{semenov21-tf} without imposing such relations locally. 
\newtext{The latter decorrelation is particularly hard to reproduce in simulations, making it a stringent test of star formation and feedback modeling \citep[][]{semenov18,fujimoto19}.}
The model was also tested in low-metallicity environments, down to 0.01 solar, typical for the early Universe \citep{polzin23,polzin24}.

The paper is organized as follows.
Section~\ref{sec:methods} briefly summarizes simulation details. Section~\ref{sec:results} presents our main results: an overview of the early turbulent evolution of our simulated MW analog and the key roles of turbulence-regulated local SFE in simultaneously producing early bursty evolution and keeping global SFEs low, which enables the survival of the early-forming disk. Section~\ref{sec:discussion}, discusses our findings and Section~\ref{sec:summary} summarizes our conclusions.

\vspace{2em}
%-----------------------------------------------------------------
%-----------------------------------------------------------------
\section{Methods}
\label{sec:methods}
%-----------------------------------------------------------------

We use a cosmological zoom-in simulation of an early-forming MW-like galaxy introduced in \citetalias{semenov24a}. The initial conditions for this galaxy were extracted from the TNG50 cosmological-volume simulation \citep{nelson19,pillepich19}
and represent one of the MW-mass disk galaxies, that resembles the MW in terms of the chemo-kinematic structure of the stellar disk, suggesting early disk spinup \citep[subfind halo ID {\tt 519311};][]{semenov23a,semenov23b,chandra23,pillepich23}. 
This galaxy is resimulated with the adaptive mesh refinement $N$-body and hydrodynamics code ART \citep{kravtsov99,kravtsov02,rudd08,gnedin11} and the ISM, star formation, and feedback model which proved remarkably successful in producing realistic SFRs and ISM properties in idealized simulations of MW-mass \citep{semenov17,semenov18,semenov19} and dwarf disk galaxies \citep[][also see the Introduction]{semenov21-tf,polzin23,polzin24}. 
The key improvements of our simulation compared to the original TNG model are summarized in Table 1 of \citetalias{semenov24a}, and the model is detailed in \citet[][]{semenov21-tf}. Here we briefly highlight its main aspects relevant to the current study. 

One of the key advantages of our model is the explicit and realistic modeling of cold ISM formation coupled with a spatially and temporally variable radiation field. 
To this end, our simulation includes explicit on-the-fly transfer of the UV radiation field modeled using the Optically Thin Variable Eddington Tensor approximation \citep[OTVET;][]{gnedin01,gnedin14}. The ionizing radiation field includes the contribution from both the stellar particles formed during the simulation and the \citet{haardt12} cosmological background.

Gas heating and cooling are modeled following \citet{gnedin12}. The metallicity-dependent part of the cooling and heating functions varies with the local radiation field, which is self-consistently modeled in the simulation. The metallicity-independent part of the cooling and heating functions is computed exactly by summing over all relevant reactions involving H and He ions and molecular hydrogen, which are modeled by solving the ``six-species'' chemical network without assuming ionization equilibrium \citep[see Appendix A.4 in][]{gnedin11}. 

Another key feature of our simulation is explicitly modeled locally variable star formation efficiency on the scales of star-forming regions, coupled with ISM turbulence. The star formation efficiency is often used to parameterize the local star formation rate density in simulations using the freefall time, $\tff = \sqrt{3\pi/32G\rho}$, as the reference timescale:
\begin{equation}
\label{eq:rhosfr}
\rhoSFR = \epsff \frac{\rho}{\tff}.
\end{equation}
Galaxy simulations often adopt thresholds---e.g., in density, temperature, etc.---to designate which gas is available for star formation and treat $\epsff$ as a tunable parameter calibrated against available observations. Such an approach necessarily relies on the assumption that the local properties of star-forming gas are universal across the environments probed in the simulation.

In our simulation, we instead allow $\epsff$ to vary continuously with the local properties of small-scale turbulence, using the fit to magnetohydrodynamic simulations of turbulent star-forming regions by \citet{padoan12}:
\begin{equation}
\label{eq:epsff-P12}
\epsff = 0.9 \exp{(-\sqrt{\avir/0.53})},
\end{equation}
where the local virial parameter is defined as for a uniform sphere with a radius equal to half of the cell size, $R = \Delta/2$ \citep{bertoldi92}:
\begin{equation}
\label{eq:avir}
    \avir \equiv \frac{5 \stot^2 R}{3GM} \approx 6 \frac{ (\stot/5\kms)^2 }{ (n/100\cc) (\Delta/25 \pc)^2},
\end{equation}
with $\stot = \sqrt{\st^2+\cs^2}$ that accounts for both the small-scale turbulent velocity dispersion, $\st = \sqrt{2\,\eturb/\rho}$, and thermal support, expressed via the sound speed $\cs = \sqrt{\gamma(\gamma-1)\,\eth/\rho}$. 

This approach relies on the assumption that the \emph{local} properties of self-gravitating turbulent medium are universal across the environments probed in the simulation, while still allowing arbitrary variations of $\epsff$ following the variations of $\avir$ set by these environments. This is a significantly weaker assumption than that star formation efficiency is universal, and it enables us to make meaningful predictions about the distribution and evolution of $\epsff$ at high redshifts. 

To compare our results with a more commonly used fixed-$\epsff$ prescription, we also resimulate our galaxy using $\epsff$ fixed at progressively higher values and adopting more stringent star formation thresholds. Specifically, we run three additional simulations with $\epsff = 0.7\%$ at $\avir < 70$, $\epsff = 10\%$ at $\avir < 10$, and $\epsff = 100\%$ at $\avir < 5$. These values are motivated by the average $\epsff$ for the corresponding $\avir$ thresholds in our fiducial simulation after the disk formation.
The results of these tests are presented in Section~\ref{sec:results:sfe-model}.

To model small-scale turbulent energy $\eturb$, that is required to estimate $\avir$, we use an explicit subgrid turbulence model following the Large-Eddy Simulation methodology \citep[see, e.g.,][for reviews]{sagaut,garnier}. Our implementation is based on the ``shear-improved’’ model of \citet{schmidt14} and detailed in \citet{semenov16}. In this model, $\eturb$ is modeled in each cell as an additional energy variable, akin to thermal energy, which is sourced by the energy cascade from the fluctuating part of the resolved velocity field and decays into heat on the timescale close to the turbulence turnover time on the scale of the cell size. Unresolved turbulence provides nonthermal pressure support and, most importantly, directly couples with the star formation prescription described above. 

Our simulations also model molecular hydrogen abundances independently from star formation, enabling us to make predictions about the relation between star formation and molecular gas in the early Universe. 
The formation and destruction of molecular hydrogen is modeled by using the ``six-species’’ nonequilibrium chemical network described in the appendix of \citet{gnedin11} that tracks the evolution of \ion{H}{1}, \ion{H}{2}, \ion{He}{1}, \ion{He}{2}, \ion{He}{3}, and H$_2$ in each cell, coupled with the local radiation field. To model H$_2$ photodissociation, the radiative transfer in the Lyman--Werner bands is modeled following \citet{ricotti02}. We also keep the modifications to the original \citet{gnedin11} model introduced in \citet{semenov21-tf}, specifically, a 100-pc ceiling on the size of the shielded regions estimated from the Sobolev approximation and reduced clumping factor of H$_2$, 3 instead of 10.

%-----------------------------------------------------------------
%-----------------------------------------------------------------
\section{Results}
\label{sec:results}
%-----------------------------------------------------------------

Here we present our main results: we outline the early evolution of the MW-like galaxy and highlight the importance of the explicit modeling of locally variable star formation efficiency, $\epsff$. In our simulation, the variations in $\epsff$ dominate the variation of the global SFR at early times, while at the same time, the low average $\epsff$ values ensure the survival of the early-forming disk. 

Our results can be interpreted as a plausible scenario for the very early evolution of the MW, even though the early stellar disk would not survive until $z=0$ in our simulation due to the destructive merger at $z\sim 2.6$. As we showed in \citetalias{semenov24a}, the early stellar disk before the merger resembles the low-metallicity stellar population in the MW, while the survival of such a disk depends on the details of ISM, star formation, and feedback modeling as well as the presence of such mergers in the mass assembly history (see also Section 4 in \citetalias{semenov24a}).

%-----------------------------------------------------------------
\subsection{Early Turbulent Evolution Overview}
%-----------------------------------------------------------------

\begin{figure*}
\centering
\includegraphics[width=0.24\textwidth]{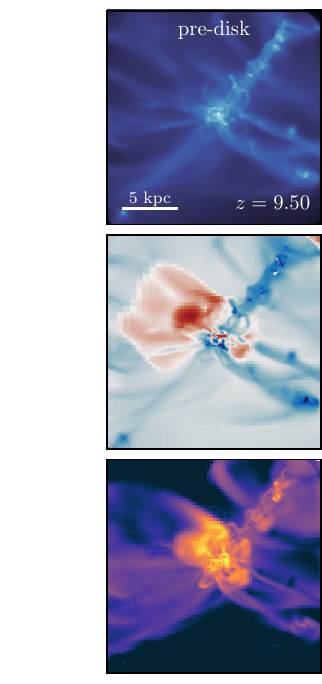}%
\includegraphics[width=0.16\textwidth]{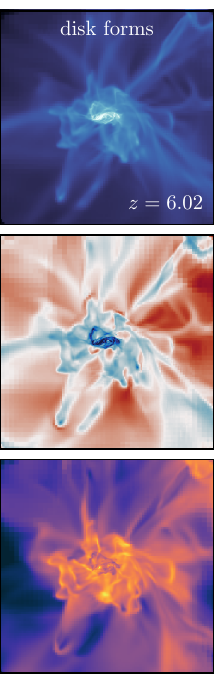}%
\includegraphics[width=0.16\textwidth]{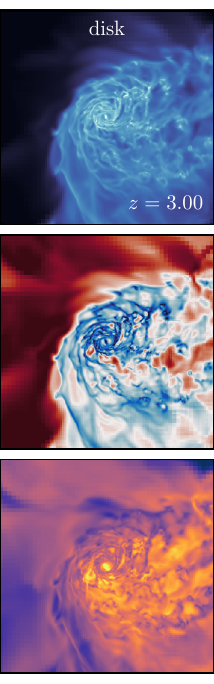}%
\includegraphics[width=0.16\textwidth]{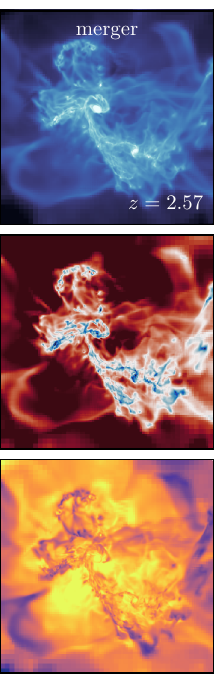}%
\includegraphics[width=0.24\textwidth]{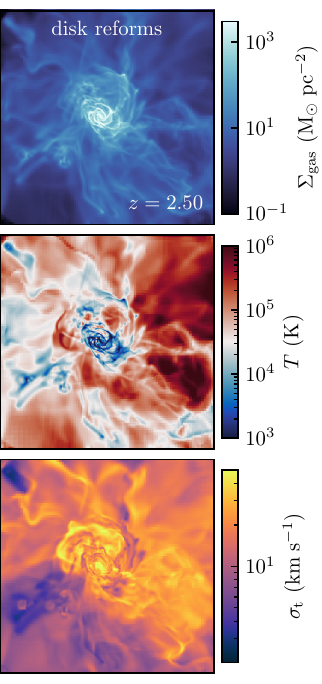}
\caption{\label{fig:maps} Projections of gas density (top), temperature (middle), and small-scale turbulent velocity (bottom) during the early evolution of our MW-like progenitor. Temperature and turbulent velocity values are mass-weighted averaged along the line of sight. The columns show the galaxy morphology at different stages of evolution. Up to disk formation, the early evolution is dominated by cold stream accretion and active mergers. A stable disk forms around $z \sim 6\text{--}7$, grows until $z\sim 2.6$, when a destructive merger occurs, after which the disk quickly reforms, by $z\sim 2.5$, and persists until the end of the simulation at $z\sim 2$. }
\end{figure*}

Figure~\ref{fig:maps} shows the early evolution of our MW progenitor. The rows show, from top to bottom, gas surface density, mass-weighted temperature, and subgrid turbulent velocity dispersion along the line of sight. 
At early times, $z > 7$, the accretion of material onto the galaxy is dominated by prominent cold streams and mergers. These streams and mergers drive turbulence in the ISM and circumgalactic medium (CGM) of the main progenitor while expanding SN bubbles start heating the gaseous halo. Around $z \sim 6\text{--}7$ a stable gaseous disk forms (see also \citetalias{semenov24a}). The disk grows until $z\sim 2.6$, when it experiences a destructive merger, but the gas disk reforms quickly after the merger, or by $z \sim 2.5$ (see also Figure~\ref{fig:jzjc-Rdisk} below). The size of the post-merger disk is significantly smaller than before the merger, leading to higher average densities of star-forming gas and higher SFRs (see Section~\ref{sec:sfh}), which in turn drive higher levels of turbulence in the CGM.

\begin{figure}
\centering
\includegraphics[width=\columnwidth]{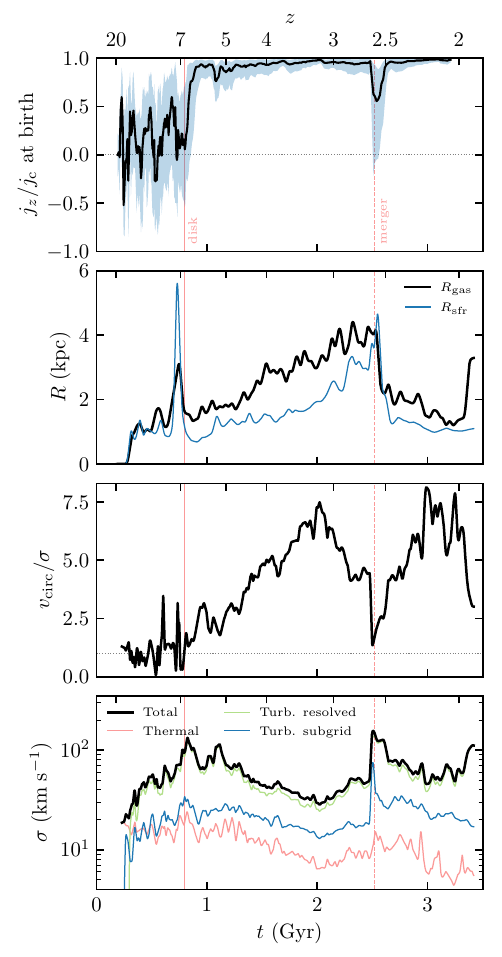}
\caption{\label{fig:jzjc-Rdisk} Evolution of key disk properties, from top to bottom: orbital circularity of young ($< 100\Myr$) stars, $\jzjc$; disk size, measured as the half-mass radius of the ISM gas with $n > 0.1\cc$ and the radius containing half of the total SFR; rotational support, defined as the ratio of $\vcirc$ at the circularization radius of the ISM gas and the total velocity dispersion of gas; and the total velocity dispersion with the contribution from resolved and unresolved turbulence and thermal motions shown separately. \newtext{The blue shaded region in the top panel indicates a 16--84 interpercentile range of $\jzjc$. To improve presentation, the values of $R$, $\vcirc/\sigma$, and $\sigma$ are smoothed on a 10 Myr timescale using a Gaussian filter.} The vertical solid and dashed red lines indicate the moments of disk formation and destructive merger, respectively. A thin ($\jzjc > 0.8$) star-forming disk spins up rapidly around $z \sim 6\text{--}7$ and grows in size and rotational support up to $z\sim 2.6$, when a destructive merger occurs. The disk reforms quickly, within $\sim 200\Myr$, but becomes significantly more compact than before the merger. Throughout the disk evolution, its velocity dispersion is dominated by resolved velocity dispersion at a $\sim 40\text{--}80\kms$ level.}
\end{figure}

The evolution of the galactic disk is quantitatively summarized in Figure~\ref{fig:jzjc-Rdisk}. The top panel shows the orbital circularities of star particles at birth, $\jzjc$, where $j_z$ are the $z$ components of the angular momentum of young ($< 100\Myr$) stellar particles and $j_{\rm c}$ are the angular momenta on circular orbits with the same total energies. At $z>7$, $\jzjc$ values exhibit a strong scatter indicative of chaotic star formation in a quasi-spheroidal configuration, while around $z\sim7$, $\jzjc$ sharply pile up at values $>0.8$, reflecting the formation of a thin star-forming disk. During the merger at $z\sim 2.6$, $\jzjc$ of young stars drop again to low values as the gaseous disk becomes disrupted, but the disk reforms shortly after, within $\sim 200\Myr$, or by $z \sim 2.5$. 

The second panel shows the evolution of the disk size, defined as either the circularization radius of the dense ($n > 0.1\cc$) gas, $\Rcirc$, or the radius containing half of the star formation. Both estimates show qualitatively similar behavior: the disk size grows monotonically until the merger, after which its radial extent drops sharply by a factor of 2--3 and stays constant until $z \sim 2$, when we stop the simulation.

The bottom two panels of Figure~\ref{fig:jzjc-Rdisk} show the rotational support of the disk, $\vcirc/\sigma$, and its velocity dispersion, with thermal, subgrid, and resolved components shown separately. The value of $\vcirc$ is calculated at the circularization radius of the dense gas, $\Rcirc$. After disk formation, the rotational support increases monotonically up to $\vcirc/\sigma \sim 7$ at $t \sim 2\Gyr$, after which it declines as the gas associated with the subsequent merger is accreted on the main progenitor driving stronger ISM turbulence $\sim 0.5\Gyr$ prior to the merger. Roughly, $\sim 200\Myr$ after the merger, $\vcirc/\sigma$ returns back to $\sim 5\text{--}7$. 

The velocity dispersion of the ISM gas is dominated by the resolved turbulence with values varying between 30--100$\kms$, with prominent peaks around the final merger before the disk formation ($z \sim 7$) and the subsequent destructive merger ($z \sim 2.6$). The subgrid turbulent velocity is subdominant to the resolved $\sigma$---as expected for a turbulent cascade containing most of the kinetic energy in the largest scales---and stays at the level of 15--30$\kms$. These values are a factor of $\sim$2--4 larger than the average thermal velocity, implying that, at the resolution scale, the gas is mostly supersonic.

%-----------------------------------------------------------------
\subsection{Star Formation History}
\label{sec:sfh}
%-----------------------------------------------------------------

\begin{figure}
\centering
\includegraphics[width=\columnwidth]{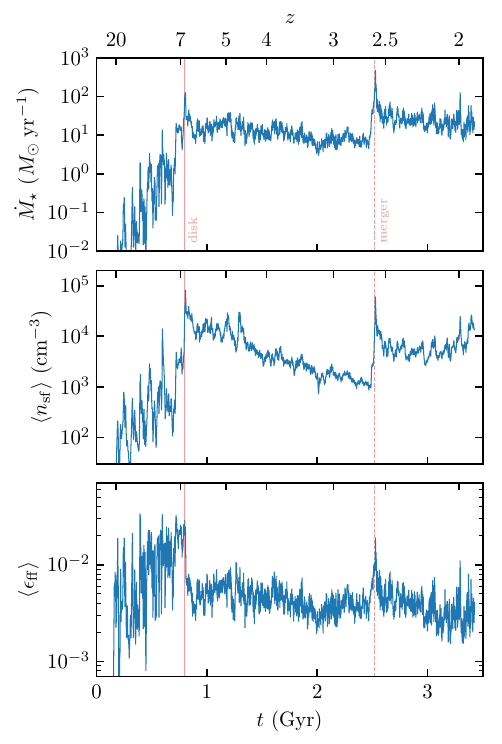}
\caption{\label{fig:sfh} Evolution of the global SFR (top), along with the key properties controlling its value: the average density of star-forming gas ($\langle \nsf \rangle$; middle) and its star formation efficiency per freefall time ($\langle \epsff \rangle$; bottom).  The vertical solid and dashed red lines indicate the moments of disk formation and destructive merger, respectively. \newtext{To highlight the variability of the presented quantities while avoiding artifacts caused by the nonuniform spacing between simulation time steps, the values are smoothed on a short, 1 Myr, timescale using a Gaussian filter.} The SFR increases exponentially before disk formation and steadily declines after, until the burst caused by the merger and subsequent decline, albeit at the level $\sim 5$ times higher than before the merger. This overall trend of SFR is set primarily by the evolution of $\langle \nsf \rangle$ and the total amount of star-forming gas (see Figure~\ref{fig:Mgas-tdep}), while $\langle \epsff \rangle$ values stay $\sim0.3\text{--}2\%$ but exhibit strong fluctuations, which translate into the short-term variability of SFR.}
\end{figure}

\begin{figure}
\centering
\includegraphics[width=\columnwidth]{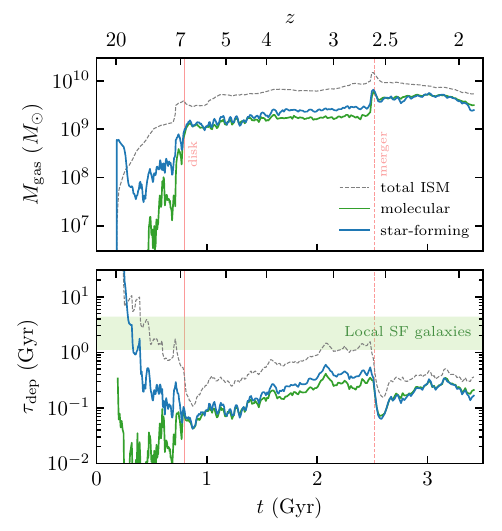}
\caption{\label{fig:Mgas-tdep} Evolution of the gas mass (top) and corresponding depletion time ($\taudep \equiv \Mgas/\SFR$; bottom). Different lines show the total ISM gas (defined as the gas denser than $n > 0.1\cc$; dashed gray), star-forming ($\epsff > 10^{-5}$; blue), and molecular gas (using local H$_2$ density predicted by the nonequilibrium chemical network and multiplied by a factor of 1.36 to account for the helium contribution; green). The vertical solid and dashed red lines indicate the moments of disk formation and destructive merger, respectively. The depletion time of total gas drops exponentially as the gas reservoir is assembled before the disk formation, after which it monotonically increases due to dense gas consumption, with a sharp drop caused by the merger. The depletion time of molecular gas after disk formation is $\sim100\text{--}300\Myr$, significantly shorter than $\sim1\text{--}4\Gyr$ derived for the nearby star-forming galaxies \citep[indicated with the green band; e.g.,][]{leroy13,bolatto17}, corresponding to a starburst regime. Interestingly, the molecular and star-forming gas are decoupled before disk formation and become strongly correlated after, as is also observed for nearby star-forming galaxies. }
\end{figure}

Figure~\ref{fig:sfh}, shows the evolution of the key properties describing the global star formation in the galaxy: the global SFR ($\SFR$; top panel), the average density of star-forming gas ($\langle \nsf \rangle$; middle panel), and the average star formation efficiency per freefall time of star-forming gas ($\langle \epsff \rangle$; bottom panel). Figure~\ref{fig:Mgas-tdep} also shows the evolution of the star-forming gas mass ($\Msf$; defined as the gas with $\epsff > 10^{-5}$)\footnote{Although the choice of the $\epsff > 10^{-5}$ cut is somewhat conservative, we find that the values of $\Msf$, $\langle \nsf \rangle$, and $\langle \epsff \rangle$ change only by a factor of $\sim 2$ if we increase this threshold to $\epsff > 10^{-3}$, which accounts for most of the star formation in the galaxy. We present the results for $\epsff > 10^{-5}$ due to the higher frequency of outputs available for this choice.} along with the mass of the total ISM (defined as gas denser than $n > 0.1\cc$) and molecular gas (modeled using the nonequilibrium chemistry network and corrected by a factor of 1.36 to account for helium). The above quantities are related via

\begin{equation}
\label{eq:sfr}
    \SFR =  \frac{\langle \epsff \rangle \Msf}{\tauff}.
\end{equation}
To ensure this correspondence, $\langle \epsff \rangle$ is weighted by $\rho^{1.5}\,dV$, while $\langle \nsf \rangle$ is the density at which the freefall time equals the mass-weighted inverse average $\tauff = \langle \tff^{-1} \rangle^{-1}$, considering only star-forming gas in both cases.

The comparison of $\SFR$, $\langle \nsf \rangle$, $\langle \epsff \rangle$, and $\Msf$ in Figures~\ref{fig:sfh} and \ref{fig:Mgas-tdep} reveals that the overall shape of $\SFR(t)$ mainly reflects the evolution of $\Msf$ and $\langle \nsf \rangle$, while the values of $\langle \epsff \rangle$ exhibit a remarkably narrow range of values.

The early rapid increase in the SFR reflects the accumulation of star-forming gas mass, $\Msf$, and the increase of its average density, $\nsf$, before the disk formation. When the disk settles, the global SFR reaches a maximum and then gradually decreases, as $\nsf$ decreases rapidly despite the increasing $\Msf$. This trend reflects the rapid depletion of the densest star-forming gas reservoir (decreasing $\nsf$) combined with the buildup of a more extended gaseous disk (increasing $\Msf$; see also the second panel in Figure~\ref{fig:jzjc-Rdisk}). The merger at $z \sim 2.6$ leads to a sharp increase of SFR and the subsequent settling at a value higher than before the merger, as it brings in new gas (increasing $\Msf$) and results in a more compact disk (and therefore higher $\nsf$).

The variation of the average $\epsff$ values is significantly smaller than these trends: $0.3\text{--}3\%$ before disk formation and $\sim 0.2\text{--}0.6\%$ after the disk formation. These values are remarkably close to the $\sim$1\% values observed in the local Universe \citep[e.g.,][]{evans09,evans14,heiderman10,lada10,lee16,leroy17,utomo18}. We stress that this outcome was not imposed in the simulation, as our star formation model allows arbitrary variations of $\epsff$. Instead, this universality of $\epsff$ emerges as a result of the efficient dispersal of low-$\avir$ star-forming regions (corresponding to high $\epsff$) by stellar feedback \citep{polzin24}. 
Similar near-universal $\epsff \sim 1\%$ values are also predicted in the VINTERGATAN zoom-in cosmological simulation of a MW-like galaxy with a different treatment of small-scale ISM turbulence by \citet{segovia-otero24}.
Note also, that even though the range of $\langle \epsff \rangle$ values is relatively small, its variation dominates the short-term variability of the global SFR (see Section~\ref{sec:results:sfr-variability}).

Interestingly, even though the average values of \emph{local} $\epsff$ are small, the \emph{global} depletion times are short owing to very high densities of star-forming gas at these early times: $\langle \nsf \rangle \sim 10^3\text{--}10^4\cc$. For example, the depletion time of molecular gas is 100--300 Myr, or a factor of 10 shorter than observed for typical local star-forming galaxies, implying that the galaxy is in the starburst regime.

The simulation also exhibits a nontrivial relation between molecular and star-forming gas reservoirs. In our simulation, the two are modeled independently: molecular gas is modeled using a nonequilibrium chemical network coupled with the self-consistent transfer of UV radiation, while star-forming gas is modeled based on the predicted distribution of turbulent energy on unresolved scales (see Section~\ref{sec:methods}). 
As Figure~\ref{fig:Mgas-tdep} shows, before the disk formation, the two are decoupled \citep[see also][]{glover12a,glover12b}. This decoupling results from the inefficient formation of molecular gas in low-metallicity environments and the rapid dispersal of molecular regions by feedback before the gas has time to become molecular \citep{polzin23}. \newtext{Star-forming gas, on the other hand, forms on local dynamical timescales, on which gas density increases and local turbulent velocity decreases, resulting in gas reaching low $\avir \propto \sigma^2/\rho$. These dynamical timescales are significantly shorter than the long timescales of molecular gas formation in low-metallicity environments. These effects} lead to significantly lower amounts and shorter depletion times of molecular gas compared to star-forming gas before the disk is in place.\footnote{\newtext{At very early times, $z \sim 20$, the amount of star-forming gas seemingly exceeds the total amount of the ISM gas in the top panel. This is an artifact of the specific choice of the thresholds used to designate the ISM ($n > 0.1\cc$) and star-forming gas ($\epsff > 10^{-5}$). At these early times, turbulent velocities did not have enough time to develop, resulting in $\epsff > 10^{-5}$ in diffuse gas. Note, however, that this diffuse gas does not contribute significantly to the total star formation due to its very long freefall times, with typical star formation proceeding at $n \sim 100\cc$ and $\epsff \sim 0.2\text{--}1\%$ (see Figure~\ref{fig:sfh}). The total amount of such actively star-forming gas is small compared to the total gas with $n > 0.1\cc$.}}

After disk formation, \newtext{the amounts of} molecular and star-forming gas become strongly correlated as is also observed for typical local star-forming galaxies, as manifested in a near-linear relation between molecular gas and SFR surface densities \citep[e.g.,][]{wong02,leroy13,bolatto17}.
Given that molecular and star-forming gas reflect distinct states of the ISM gas, this correlation implies that the gas distribution evolves near-self-similarly, such that the \newtext{total} fractions of gas in the molecular (i.e., dense and self-shielded) and star-forming (i.e., low $\avir$) states change synchronously, \newtext{even though these phases do not necessarily coincide spatially (see Section~\ref{sec:discussion:H2} for further discussion). Also, note that the fact that the amounts of molecular and star-forming gas in Figure~\ref{fig:Mgas-tdep} coincide within a factor of $\sim$2 should be interpreted with caution as the total star-forming mass is sensitive to the choice of the $\epsff$ threshold that defines this gas ($\epsff > 10^{-5}$ in our case).}

Finally, it is also worth noting that the global depletion time of total ISM gas experiences a sharp decline before disk formation, and saturation with a mild increase after \citep[see also][]{vintergatan4}. Such behavior is qualitatively similar to the picture proposed by \citet{conroy22} to explain the surprising increase of $\alpha$-element abundances in low-metallicity MW stars with [Fe/H] $< -1.5$. Although in our simulation, the values of depletion times are substantially different from those reported by these authors (see Section~\ref{sec:discussion:regimes}), our results provide a plausible explanation for such a drastic change in the global star formation efficiency (the inverse depletion time in \citealt{conroy22}). The early inefficient stage reflects the buildup of the early star-forming reservoir, while the late efficient stage reflects star formation in a galactic disk.

%-----------------------------------------------------------------
\subsection{The Turbulent Origin of SFR Variability}
\label{sec:results:sfr-variability}
%-----------------------------------------------------------------

\begin{figure}
\centering
\includegraphics[width=\columnwidth]{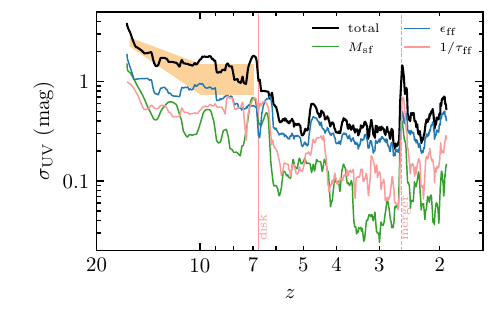}
\caption{\label{fig:sfr-scatter} The evolution of the UV magnitude variability (see Equation~\ref{eq:sigmaUV}) together with a contribution from the terms in Equation~\ref{eq:sfr}: total mass of star-forming gas ($\Msf$), its average freefall time ($1/\tauff \equiv \langle 1/\tff \rangle$), and variable star formation efficiency per freefall time ($\langle \epsff \rangle$). The evolution is shown in $\log\,z$ to visually stretch the pre-disk stage, where the shaded orange region shows the values of $\sigma_{\rm UV}$ that can explain high-redshift JWST UV-luminosity functions as reported by \citet[][]{shen23}. The vertical solid and dashed red lines indicate the moments of disk formation and destructive merger, respectively. The variation of SFR is dominated by the variation of $\langle \epsff \rangle$ at all times except close to the disk formation and merger, when all three quantities contribute comparably due to the large jumps in $\Msf$ and $\langle \nsf \rangle$ (recall Figures~\ref{fig:sfh} and \ref{fig:Mgas-tdep}).}
\end{figure}

One of the key effects of the explicit modeling of cold turbulent ISM is the significant increase in the burstiness of star formation compared to the model based on the effective equation of state approach in the original TNG50 run (see \citetalias{semenov24a}). This is particularly enticing in the context of JWST observations of the large abundances of early galaxies, which can be explained by large variations of SFRs at early times (see references in the Introduction). In this section, we investigate the origin of the SFR variation in our simulation.

Generally, the variation of the global SFR can be caused by variations in either of the terms in Equation~(\ref{eq:sfr}): the total amount gas available for star formation ($\Msf$), its average density, or freefall time ($\tauff$), and its average star formation efficiency per freefall time, reflecting the turbulent state of star-forming gas ($\langle \epsff \rangle$). Figure~\ref{fig:sfr-scatter} shows the variability of the total SFR, expressed in terms of the scatter in the UV absolute magnitude, $\sigma_{\rm UV}$, and the contributions from these three terms separately. To convert the SFR scatter to $\sigma_{\rm UV}$, we use Equation (10) from \citet{madau14}, which implies that the bolometric UV luminosity is proportional to the SFR and therefore, by definition, 
\begin{equation}
\label{eq:sigmaUV}
    \sigma_{\rm UV} \approx 2.5\;\sigma(\log_{10}\SFR),
\end{equation}
where we calculate $\sigma(\log_{10}\SFR)$ over a running 100\,Myr window to estimate the temporal variation of the SFR variability. The SFR and all other terms are sampled at the global step cadence of our simulation, $\sim 0.5$ Myr timescale.

The figure shows that the UV variability is high early on and decreases with time. This trend, and the values of $\sigma_{\rm UV} \sim 1\text{--}2$ mag at $z > 7$, are consistent with the variability that is required to explain the large abundance of early UV-bright galaxies discovered by JWST as reported by \citet[][]{shen23}: $\sigma_{\rm UV} \sim 2.5$ at $z \sim 16$, and decreasing to $\sigma_{\rm UV} \sim  0.75\text{--}1.5$ at $z \sim 7\text{--}10$ (see the orange shaded region). At later times, after the disk formation, $\sigma_{\rm UV}$ continues to decrease (see also Figure~\ref{fig:sfh}).

As Figure~\ref{fig:sfr-scatter} also shows, the variation of SFR is almost always dominated by the variation in $\epsff$ (the blue line). The only exceptions when the variations in all three terms contribute comparably are the moment of disk formation (associated with a merger) and the later merger at $z \sim 2.6$, as $\Msf$ and $\tauff$ experience sudden jumps contributing to their dispersion on a 100 Myr timescale. This is consistent with the visual impression from the evolution of these different terms in Figures~\ref{fig:sfh} and \ref{fig:Mgas-tdep}. Indeed, while the long-term evolution of SFR reflects that of $\Msf$ and $\tauff$, the short-term (few Myr) variation is dominated by $\epsff$ variations. 

\begin{figure}
\centering
\includegraphics[width=\columnwidth]{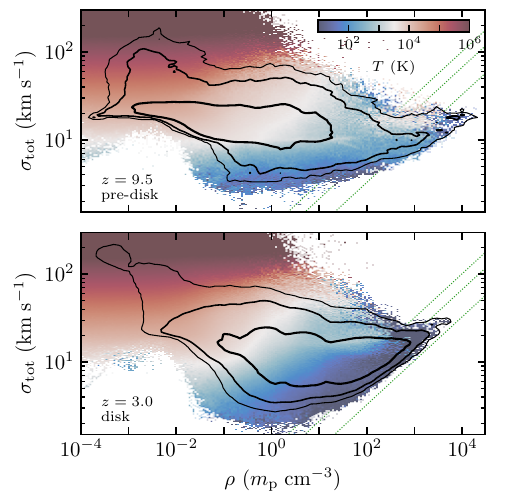}
\caption{\label{fig:nsigma} The distribution of gas in the $n$--$\stot$ space before (top) and after disk formation (bottom). The local $\stot = \sqrt{\st^2 + \cs^2}$ includes the contributions from unresolved turbulence and thermal pressure. The diagonal dotted lines show constant values of $\avir$ corresponding to, from top to bottom, $\epsff = 0.1\%$, 1\%, and 10\%. The difference in the average $\epsff$ and its variability before and after disk formation (recall the bottom panel of Figure~\ref{fig:sfh}) is caused by the difference in the gas distribution in the high-$\epsff$ regime, which in turn is caused by the different regimes of star-forming gas formation: stirring and compression by accretion filaments and mergers at early times versus ISM turbulence decay and disk instabilities at late times (see the text). }
\end{figure}

The difference in the UV and SFR variability before and after the disk formation can be explained by the difference in the small-scale turbulent support of gas. Figure~\ref{fig:nsigma} shows the distribution of gas in the $n$--$\stot$ plane, the location in which determines the virial parameter of gas, $\avir \propto \stot^2/n$, and therefore its local $\epsff$ (see Section~\ref{sec:methods}). The diagonal dotted lines show constant values of $\avir$ corresponding to, from top to bottom, $\epsff = 0.1\%$, 1\%, and 10\%. 

The variations in the gas distribution in the high-$\epsff$ part of the $n$--$\stot$ plane result in the variation of the average $\langle \epsff \rangle$.
Before disk formation (top panel) most of the gas resides at low $n < 10\cc$ with only the high-density tail of the distribution probing the star-forming regime, leading to large chaotic variations of $\langle \epsff \rangle$. In contrast, when the disk is formed (lower panel), a significant fraction of gas piles up along these high-$\epsff$ values. This pileup reflects the continuous formation and dispersal of star-forming gas in the disk ISM \citep[see also][]{semenov17,semenov18}. As a significantly larger fraction of gas participates in star formation, the variations of $\langle \epsff \rangle$ are statistically averaged out and decrease in magnitude.

Apart from the level of $\langle \epsff \rangle$ variation, the behavior of the PDF in Figure~\ref{fig:nsigma} also explains the offset in normalization of $\epsff$ before and after disk formation (recall Figure~\ref{fig:sfh}). At early times, the high-density tail probes deeper into the low-$\avir$ (high $\epsff$) regime, reaching values significantly higher than 10\%, while after disk formation, the highest local $\epsff$ only reach $\lesssim 10\%$.
On average, this leads to a mild change of $\langle \epsff \rangle$, even though locally, the highest $\epsff$ values differ substantially (see also Sections~\ref{sec:results:sfe-model} and \ref{sec:discussion:regimes}).

Physically, the difference in the gas distributions in Figure~\ref{fig:nsigma} stems from the difference in the mode of the formation of actively star-forming gas. 
Before the disk is in place, the formation of dense star-forming gas is driven by the active accretion of gas in cold filaments and frequent mergers (recall the first two columns in Figure~\ref{fig:maps}). These violent processes drive strong ISM turbulence and push the gas into the low-$\avir$ regime. The highly episodic nature of these processes also leads to large variations of $\langle \epsff \rangle$.

After the disk forms, in contrast, the formation of dense gas ISM is driven by the local dissipation of turbulence, gas compression in spiral arms and colliding SNe shells, and other types of local disk instabilities. These processes are milder and can only sustain star-forming gas with relatively low, sub-10\%, $\epsff$ values, leading to lower $\langle \epsff \rangle$ than before the disk. As star-forming gas formation becomes distributed throughout the ISM, this also leads to smaller fluctuations of $\langle \epsff \rangle$.

Overall, the variation in $\epsff$ reflects the rapid stochastic changes of the ISM turbulence state in the early galaxy driven by different processes before and after disk formation. Such a variation is usually not accounted for in galaxy simulations that commonly adopt fixed $\epsff$ and treat it as a tunable parameter. Therefore, the variability of SFR in such simulations can be underestimated, which can be a reason why such simulations might underpredict the abundances of early UV-bright galaxies revealed by JWST.
As we demonstrate in the next section, the SFR variability can be increased by choosing a high fixed $\epsff$ value, which, however, can damage the early disk and even prevent it from forming.

%-----------------------------------------------------------------
\subsection{Effect of Star Formation Modeling}
\label{sec:results:sfe-model}
%-----------------------------------------------------------------

\begin{figure*}
\centering
\includegraphics[width=\textwidth]{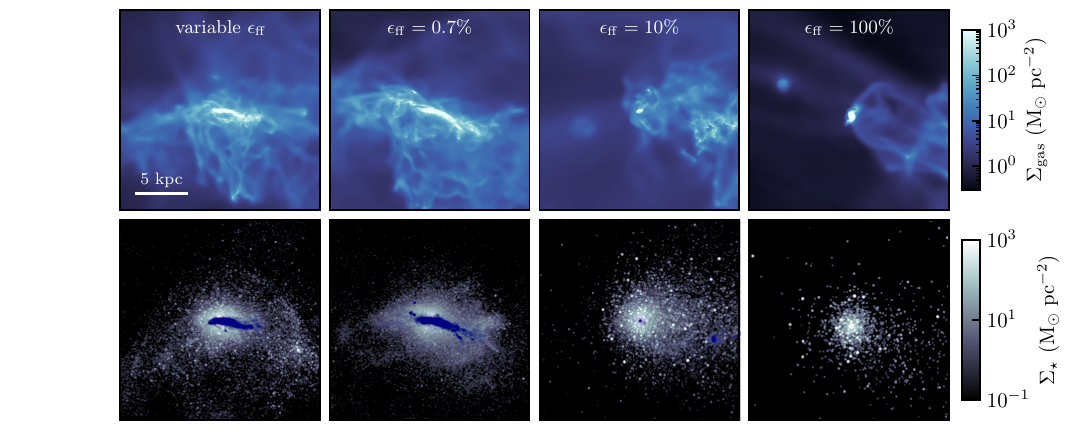}
\caption{\label{fig:maps-vareff} Edge-on projections of gas (top) and stellar surface densities (bottom) in our resimulations with different star formation prescriptions at $z\approx4.5$. The blue color indicates the locations of young, $< 100\Myr$, stellar particles. For presentation purposes, the stellar surface density is smoothed using a 100 pc-wide 2D Gaussian filter to make individual massive stellar particles more apparent. Our fiducial run with variable $\epsff$ and low fixed $\epsff = 0.7\%$ values exhibit qualitatively similar morphologies with an extended stable star-forming gas disk. High-$\epsff$ runs, in contrast, are qualitatively different: the gas distribution is strongly concentrated in the center, while the stellar population does not show a clear young disk and has a significant fraction of mass in massive stellar particles (bright spots in the bottom maps).}
\end{figure*}

\begin{figure}
\centering
\includegraphics[width=\columnwidth]{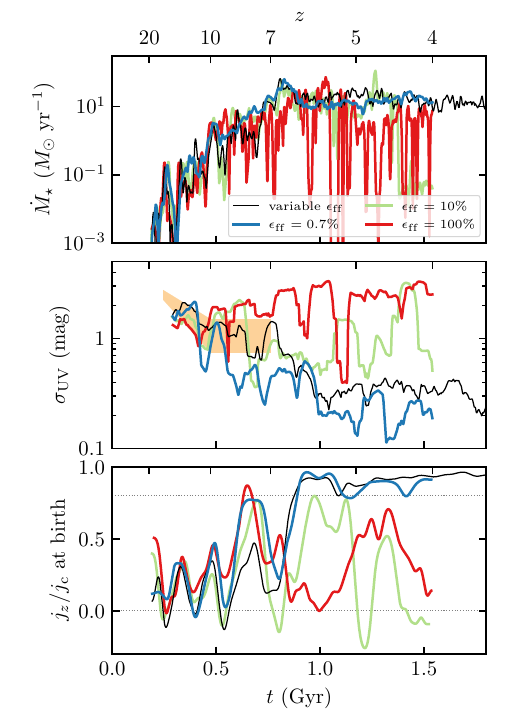}
\caption{\label{fig:sfh-vareff} Comparison of global SFR values (top), UV magnitude variability (middle; see Equation~\ref{eq:sigmaUV}), and the median circularity of the young, $< 100\Myr$, stellar population ($\jzjc$; bottom) in our resimulations with different star formation prescriptions. The shaded orange region in the second panel shows the values of $\sigma_{\rm UV}$ that can explain high-redshift JWST UV-luminosity functions as reported by \citet[][]{shen23}. The horizontal dotted lines in the bottom panel indicate the values of $\jzjc = 0$ and 0.8, corresponding to purely spheroidal and disk distributions. \newtext{To simplify the comparison between different curves in the bottom panel, the values of $\jzjc$ are smoothed on a 30 Myr timescale using a Gaussian filter.} The low $\epsff = 0.7\%$ run shows the evolution of average SFR and $\jzjc$ that is qualitatively similar to our fiducial variable $\epsff$ run, while the SFR scatter is a factor of 3 lower. High-$\epsff$ runs show substantially larger SFR values and scatter and different behavior of the star-forming disk: the disk is substantially thicker in the $\epsff=10\%$ run ($\jzjc \sim $ 0.3--0.8) and is completely missing in the $\epsff=10\%$ run ($\jzjc \sim -0.3\text{--}0.5$).}
\end{figure}

The above results show that explicit modeling of locally variable star formation efficiency per freefall time, $\epsff$, can produce highly variable global SFRs and enable early formation and survival of the stellar disk. In this section, we investigate the effect of the choice of the star formation prescription on these results. To this end, we resimulate our galaxy keeping all the parameters the same but using a fixed value of $\epsff$ below a fixed threshold in the local virial parameter, $\avir$, instead of allowing $\epsff$ to vary continuously with $\avir$ (see Section~\ref{sec:methods}). Specifically, we investigate three models: a low-efficiency model with $\epsff = 0.7\%$ in gas with $\avir < 70$, which enables star-formation throughout most of the cold ISM with the SFE close to the average value in our fiducial run (recall Figure~\ref{fig:sfh}); and two high-efficiency models with $\epsff = 10\%$ at $\avir < 10$ and $\epsff = 100\%$ at $\avir < 5$. The latter values of $\epsff$ and $\avir$ thresholds are motivated by the typical SFEs reached in low-$\avir$ gas in our fiducial simulation. 

Figure~\ref{fig:maps-vareff} compares the projections of gas and stars at $z \approx 4.5$. By that time, an extended star-forming disk forms only in the runs with low SFE: our fiducial run with variable $\epsff$ and that with fixed $\epsff = 0.7\%$. In contrast, in the runs with high SFE ($\epsff = 10\%$ and 100\%), the gas distribution is strongly centrally concentrated, while the stellar distribution is near-spherical. This difference is a combined effect of feedback kicking in only when gas reaches the low-$\avir$ state and faster dispersal rates of star-forming gas with increasing $\epsff$ \citep{semenov17,semenov18,semenov21-tf}.

In the low-SFE runs, as the gas evolves toward the low-$\avir$ state as a result of compression and turbulence dissipation, the local SFR gradually ramps up due to both the density-dependence of $\rhoSFR \propto \rho^{1.5}$ and, in our fiducial run, the exponential increase of $\epsff$ at low $\avir$ (see Equations~\ref{eq:rhosfr} and \ref{eq:epsff-P12}). As a result, the SFR becomes more uniformly distributed throughout the ISM (see the blue color in the bottom panels of the figure), with some of the high-$\avir$ regions dispersed before they become too dense and actively star-forming. Such distributed injection of feedback energy and momentum leads to relatively mild disturbances of the ISM allowing the gas disk to survive.

In the high-SFE runs, in contrast, stars are not forming until the gas reaches the low-$\avir$ state. At this point, the gas is already dense and actively star-forming leading to rapid conversion into stars and quick dispersal of remaining gas by feedback. This rapid dispersal is imprinted in the distribution of stars: (i) the instantaneous fraction of star-forming gas and therefore young stars becomes significantly smaller (see the blue color in the bottom panels; also, \citealt{semenov18,semenov21-tf}) and (ii) the stellar mass becomes ``locked'' in significantly more massive stellar particles visible as bright spots in stellar surface density maps in the bottom row of the figure. The latter is a direct consequence of significantly shorter lifetimes of star-forming regions at high $\epsff$, and can in principle produce testable differences in the populations of globular clusters, which could be remnants of stars formed in highly active star-forming environments at early times (see Section~\ref{sec:discussion:sfe-modeling} for discussion). In this process, only the gas that makes it to the central region---where the potential well is too deep for feedback to efficiently disperse the gas---can survive leading to the formation of a strongly centrally concentrated distribution.

The effects of SFE modeling on the SFR history and disk formation are further quantified in Figure~\ref{fig:sfh-vareff}. 
The SFRs shown in the figure are calculated using the stars younger than 10 Myr, as opposed to the instantaneous SFR in Figure~\ref{fig:sfh}. This is because, in the runs with high $\epsff$, the instantaneous SFR computed by integrating Equation~(\ref{eq:rhosfr}) is biased high as the actual SFR becomes limited by the amount of gas locally available for star formation and the very short lifetime of that gas \citep[see also][]{semenov17}. In the runs with low $\epsff$ (our fiducial and $\epsff=0.7\%$) both estimates of SFR produce qualitatively similar results, but for consistent comparison, here we switch to the 10\,Myr averaged SFR for all four runs.

The $\epsff = 0.7\%$ run on average follows the fiducial run in terms of the global SFR magnitude (the top panel of Figure~\ref{fig:sfh-vareff}) and the timing of disk formation (bottom panel). The SFR scatter (expressed via the UV magnitude variability $\sigma_{\rm UV}$; see Equation~\ref{eq:sigmaUV}), however, is noticeably lower (by a factor of $\sim$3 from the middle panel). This is because the SFR scatter in our fiducial run is dominated by the local variability of $\epsff$ (see Section~\ref{sec:results:sfr-variability}), the process that is inherently missing in fixed-$\epsff$ runs.

The high-$\epsff$ runs, in contrast, produce qualitatively different results. The global SFR values and scatter are similar to those in the fiducial run at very early times: $z > 7$ and $z>10$ for the $\epsff = 100\%$ and 10\%, respectively. This is because the early star formation in our fiducial run primarily occurs in highly efficient but sporadically forming regions (see Section~\ref{sec:results:sfr-variability} and Figure~\ref{fig:nsigma}). At later times, however, the high-$\epsff$ runs exhibit orders of magnitude larger variability of global SFRs. Interestingly, the average SFRs are systematically lower suggesting that stronger clustering of star formation and feedback leads to more efficient removal of gas from the galaxy \citep[see also][]{smith21,keller22}. As described above, this is because feedback is injected only in dense, low-$\avir$ star-forming regions primarily concentrated in the galaxy center, making it hard for feedback to efficiently disperse star-forming gas and stop star formation locally. The episodes of accumulation of star-forming gas and strong bursts of SFR cause strong variations in the total mass and density of the star-forming gas, which results in the orders of magnitude variation of the global SFR and UV magnitude, $\sigma_{\rm UV}$.
These vigorous variations of the SFR do not allow the extended star-forming gas disk to settle, with most of the gas accumulating in the very inner part of the galaxy. As the bottom panel of Figure~\ref{fig:sfh-vareff} shows, this leads to the absence of a clear disk in the $\epsff = 10\%$ and $100\%$ runs.

To sum up, the modeling of star formation produces substantial, competing effects on the SFR variability and disk formation. Low $\epsff$ values distributed uniformly throughout the ISM lead to smoother SFR histories and result in a milder effect of stellar feedback on the gas enabling disk survival. High $\epsff$ concentrated in the densest, lowest $\avir$ gas, in contrast, leads to significantly burstier SFR and vigorous feedback injection resulting in disk dispersal at early times. Simulation with locally variable $\epsff$ transitions from high- to low-$\epsff$ regime around the moment of disk formation enabling both highly variable SFR at early time and disk survival afterward.

%-----------------------------------------------------------------
%-----------------------------------------------------------------
\section{Discussion}
\label{sec:discussion} 
%-----------------------------------------------------------------

\begin{figure*}
\centering
\vspace{1em}
\includegraphics[width=0.9\textwidth]{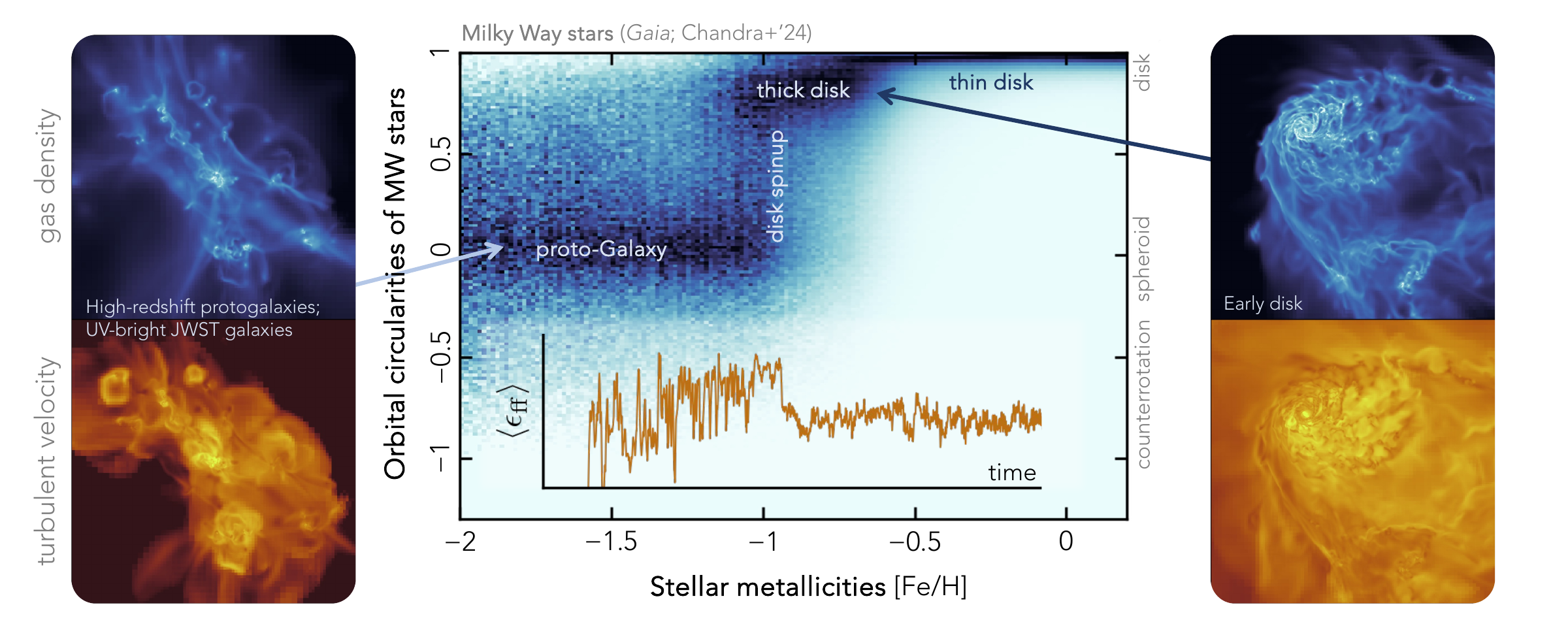}
\caption{\label{fig:sketch1} Imprints of the early galaxy evolution in the chemo-kinematic structure of the MW disk. The blue histogram shows the distribution of orbital circularities, $\jzjc$, and metallicities of the MW's stars from Figure~5 in \citep{chandra23}. The low-metallicity ([Fe/H] $< -1$) kinematically hot proto-Galaxy \citep[or ``Aurora'' in][]{bk22} reflects the early pre-disk stage when, in our simulation, vigorous variations of the ISM turbulence led to a large scatter of $\epsff$ and UV luminosities, analogous to the UV-bright early galaxies observed by JWST at $z \gtrsim 10$ (see also \citetalias{semenov24a}). The sharp transition to the disk configuration (i.e., the spinup at $-1.2 \lesssim$ [Fe/H] $\lesssim -0.9$) reflects the formation of a stable star-forming disk, after which the mode of turbulence driving switches to a new regime with smaller $\epsff$ variations, which enables the survival of the early disk, observed as the ``thick disk'' population in the present-day MW.}
\end{figure*}

\begin{figure}
\centering
\vspace{1em}
\includegraphics[width=\columnwidth]{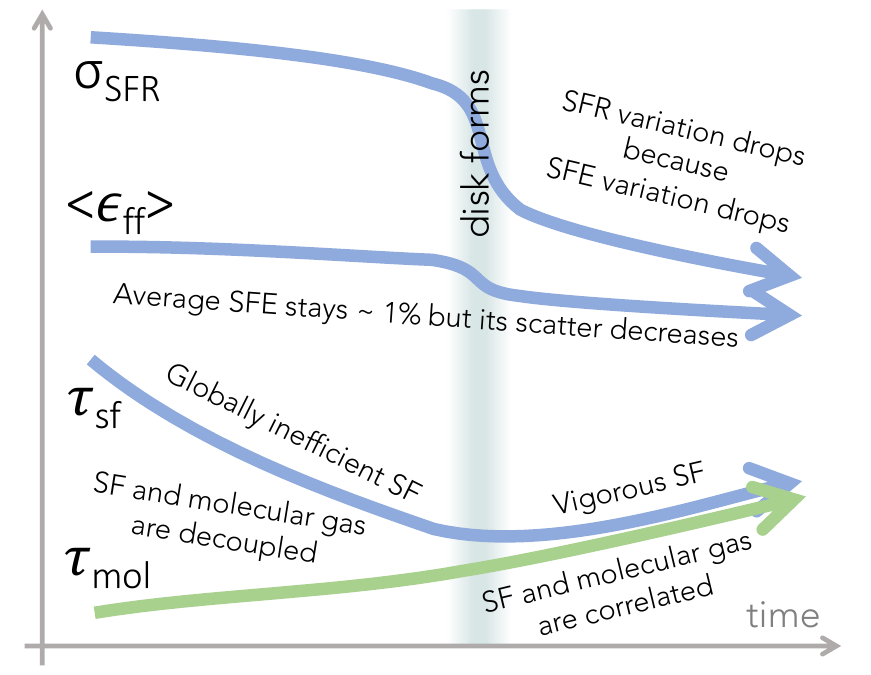}
\caption{\label{fig:sketch2} A qualitative sketch of the transition between the regimes before and after disk formation. The arrows show the qualitative evolution in the global SFR variability ($\sigma_{\rm SFR}$), average star formation efficiency (SFE) per freefall time ($\epsff$), and global depletion times of star-forming ($\tau_{\rm sf} \equiv \Msf/\SFR$) and molecular gas ($\tau_{\rm mol} \equiv M_{\rm mol}/\SFR$). Although, on average, $\epsff$ stays near-universal at $\sim 1\%$, its scatter drops significantly after disk formation, leading to the transition from highly bursty early evolution with high local $\epsff$ to the formation of a stable star-forming disk with low local $\epsff$, enabling disk survival. }
\end{figure}

Here we summarize the two regimes of galaxy evolution before and after the formation of a stable star-forming gas disk and associated qualitative change in the turbulence driving and SFR variability. These stages can be mapped onto the chemo-kinematic populations of MW stars, reflecting the formation and evolution of the MW's disk (see Figure~\ref{fig:sketch1} and \citealt{bk22,bk23,conroy22,rix22,chandra23}). Figure \ref{fig:sketch2} provides a visual summary of the changes in star-forming properties of the galaxy as a result of this transition.

%-----------------------------------------------------------------
\subsection{Transition from High to Low $\epsff$ before and after Disk Formation}
\label{sec:discussion:regimes}
%-----------------------------------------------------------------

Our simulation predicts two qualitatively different regimes of galaxy evolution before and after disk formation (see Figures~\ref{fig:sketch1} and \ref{fig:sketch2} for a visual summary). Before disk formation, the galaxy mass assembly is dominated by frequent mergers and violent accretion of gas along dense cold filaments. During this epoch, the SFR exponentially increases as the star-forming gas reservoir is accumulated and its average densities rapidly grow. The local star formation efficiency of gas during this epoch is rather low, $\sim 1\%$, however, some of the regions can be pushed to the low-$\avir$ regime reaching high $\epsff > 10\%$ values. The average $\epsff$ also experiences significant fluctuations as only a small fraction of total halo gas is actively star-forming before it is dispersed by feedback. These fluctuations contribute significantly to the variation of the global SFR during this early phase. Most of the star formation occurs in primarily neutral gas as it does not have enough time to become fully molecular before dispersal by feedback \citep[see also][]{polzin23}.
In the present-day MW stars, this stage is imprinted in the low-metallicity ([Fe/H] $< -1$) kinematically hot proto-Galaxy \citep[or ``Aurora'' in][see Figure~\ref{fig:sketch1}; also, \citealt{conroy22,rix22}]{bk22}.

After the disk forms, star formation becomes substantially less violent. The global SFR gradually decreases as the densest star-forming gas is depleted, while the total gas mass continues to increase as the disk outskirts---and therefore its size---grows. The values of $\epsff$ reached by star-forming gas drop below $\epsff < 10\%$ as the disk instabilities and decay of ISM turbulence are not able to push the gas to as low an $\avir$ regime as the violent processes before the disk formation. 
Correspondingly, the average local $\epsff$ decreases by a factor of $\sim$3, and its variations become milder in comparison to the pre-disk stage due to statistical averaging over a significant gas mass in the star-forming state. Nevertheless, these variations of $\langle \epsff \rangle$ dominate the short-term variation of the global SFR at this stage (Figure~\ref{fig:sfr-scatter}). Finally, star-forming and molecular gas become correlated, even though they reflect distinct states of the ISM gas, implying that the amount of gas in the self-shielded and low-$\avir$ states evolve synchronously (Figure~\ref{fig:Mgas-tdep}).
In the present-day MW stars, this early disk is reflected in the ``thick disk'' population (Figure~\ref{fig:sketch1}).

Qualitatively, these differences indicate that as the disk forms and the mode of star-forming gas supply changes from violent accretion in the cold stream and mergers to more mild disk instabilities and ISM turbulence decay, the galaxy transitions from high to low $\epsff$ regime described in Section~\ref{sec:results:sfe-model} \citep[see also][]{segovia-otero24}. The early bursty evolution is analogous to the runs with $\epsff$ fixed at high values, 10\%--100\%, while later mild evolution is reminiscent of low-$\epsff$ runs. By design, such a transition cannot be captured in simulations with fixed $\epsff$ values, which can lead to the trade-off between early bursty SFR and survival of early disks.

The global depletion time of gas also exhibits different behavior before and after disk formation (see the sketch in Figure~\ref{fig:sketch2}). Before disk formation, gas depletion times are long and exponentially decrease as the star-forming reservoir is built up. After disk formation, the depletion time reaches rather short values, an order of magnitude shorter than observed in the local star-forming galaxies, indicating a starburst regime. This behavior---early inefficient and later efficient global star formation---is qualitatively similar to the picture proposed by \citet{conroy22} to explain the surprising increase of $\alpha$-element abundances in low-metallicity stars with [Fe/H] $< -1.5$ without clear evidence for flattening. 
Note, however, that the values of the global depletion times, $\taudep$, reported by \citet{conroy22} are significantly longer compared to the values that we find in our simulation: $\taudep \sim 50\Gyr$ at the early inefficient ``simmering'' stage and $\taudep \sim 2.5\Gyr$ (typically for the present-day star-forming galaxies) at the later efficient ``boiling'' stage. Our simulation, in contrast, predicts $\taudep \sim 1\text{--}5\Gyr$ before the disk formation, and $\taudep \sim 0.1\text{--}0.5\Gyr$ right after (Figure~\ref{fig:Mgas-tdep}). 
Nevertheless, our simulations provide a plausible mechanism for such a transition between the ``simmering'' and ``boiling'' stages as a result of disk formation. The quantitative effect of this transition on the $\alpha$-element abundances of stars in simulations requires further investigation.

%-----------------------------------------------------------------
\subsection{Effect of Star Formation Modeling on SFR Variability and Disk Survival}
\label{sec:discussion:sfe-modeling}
%-----------------------------------------------------------------

The results presented here and in \citetalias{semenov24a} demonstrate that the early galaxy evolution and disk formation are highly sensitive to modeling of cold turbulent ISM, star formation, and feedback. In particular, the modeling of locally variable star formation efficiency per freefall time coupled with the ISM dynamics has a significant impact.

Indeed, as our results show, simulations with fixed high values of SFE per freefall time ($\epsff = $ 10\%--100\%) lead to highly variable SFR. 
This is because all of the SFR becomes concentrated in a small fraction of star-forming gas resulting in significant stochastic fluctuations of the total mass and average densities of such gas \citep{semenov17,semenov18}. In addition, with such star formation prescriptions, the gas can evolve to the star-forming state unaffected by prompt feedback until it reaches that state, making the effect of feedback more clustered and damaging for the ISM gas reservoir \citep[see also][]{smith21,keller22}.
Vigorous episodes of feedback injection associated with such high SFR variability and clustering can lead to the dispersal of early star-forming disks thereby delaying disk formation. 
In contrast, low $\epsff$ values distributed throughout the ISM lead to significantly smoother SFR histories and milder feedback effects on the ISM, enabling the survival of the early-forming disk.

Models with fixed $\epsff$ also introduce a nontrivial dependence on the thresholds used to designate star-forming gas. More stringent star formation thresholds can result in more variable SFRs \citep[e.g.,][]{buck19}, as a smaller amount of gas can be star-forming at any given moment due to a shorter lifetime and longer formation time of such gas \citep{semenov18}. In addition, if the adopted thresholds are too stringent, gas may not become star-forming before it forms clumps that are too dense for feedback to disperse. An example of such a behavior are the dense gas concentrations in the galaxy center in our simulations with high fixed $\epsff = 10\%$ and 100\% in Figure~\ref{fig:maps-vareff}.

Our fiducial model, where $\epsff$ varies continuously with the local value of the subgrid virial parameter, does not require any ad hoc thresholds and it switches between the high and low $\epsff$ regimes around the moment of disk formation (Section~\ref{sec:discussion:regimes}). At early times, only a small fraction of gas can reach the star-forming state but the gas is driven into this state by violent processes, such as active accretion and mergers, it can reach low $\avir$ and high $\epsff > 10\%$ values. When the disk settles, the average $\epsff$ and its variability also decrease. As a result, the variability of global SFR also decreases and star formation becomes more distributed across the ISM enabling the survival of the galactic disk.

In our simulation, this transition between high ($>10\%$) and low ($<10\%$) local $\epsff$ values is driven by the decrease in the $\epsff$ variability as the average $\langle \epsff \rangle$ stays relatively low ($\lesssim 1\%$; see Figure~\ref{fig:sfh}). As $\epsff$ variations dominate the global SFR variation in our simulation (see Figure~\ref{fig:sfr-scatter}), the variability of SFR also decreases after disk formation. Previous works based on cosmological simulations with explicit modeling of cold ISM have also reported a coincidence between gas disk settling and a reduction in the SFR variability \citep[e.g.,][based on the FIRE simulation suite]{stern21,yu21,yu22,hafen22,gurvich22}. The physical cause of this transition is still debated, with the proposed scenarios including the formation of the hot thermally supported inner gas halo \citep{stern19,stern20,stern21}, changes in the galactic gravitational potential \citep[e.g.,][]{hopkins23disk}, and variations in the properties of recreation flows and mergers \citep[e.g.,][]{dekel20}.
Nevertheless, it is interesting that these simulations generally predict highly bursty early SFR and relatively late disk formation \citep[][]{bk22,mccluskey23}, which is consistent with our results for the simulations with local $\epsff$ fixed at high values. 

This interplay between star formation variability and disk survival implies that considering both simultaneously is a stringent test of star formation and feedback modeling. A realistic star formation and feedback model must reproduce both the abundance of UV-bright galaxies at very early times ($z \gtrsim 10$) and the statistical occurrence of disks at later times \citep[$z \lesssim 8$; e.g.,][]{ferreira22b}. We leave a detailed comparison with these observations to future work, as it requires exploring a representative sample of zoom-in simulations with different ICs or applying the model in large cosmological volumes. 

As the results of Section~\ref{sec:results:sfe-model} suggest, another possible test of star formation and feedback modeling is the statistics of globular clusters, which can be remnants of star formation in highly active star-forming environments at early times. Indeed, cosmological simulations with higher local $\epsff$ and explicit modeling of continuous star cluster formation tend to produce steeper cluster mass functions due to earlier dispersal of star-forming gas \citep[e.g.,][]{li17a}. The transition to the low $\epsff$ regime after disk formation implies that globular clusters preferentially form before the disk formation or during the episodes of mergers, which can drive the ISM gas back to the high-$\epsff$ regime \citep[e.g.,][]{renaud15}.

%-----------------------------------------------------------------
\subsection{Note on Modeling Star Formation Based on Molecular Gas}
\label{sec:discussion:H2}
%-----------------------------------------------------------------

Our simulations show that molecular and star-forming gas phases are decoupled during the pre-disk stage in the early universe. This is not surprising given that they reflect different states of the ISM gas that do not necessarily coincide spatially: dense self-shielded and low-$\avir$ states. The decoupling in the early galaxies is mainly due to the long formation times of molecular gas in the low-metallicity environment and its fast dispersal by feedback leading to the low overall molecular abundances, implying that on tens of parsec scales---the resolution of our simulation---star formation proceeds in neutral gas before the region can become fully molecular \citep[see also][]{glover12a,glover12b,polzin23}. Note that at high metallicities in the present Universe, molecular and star-forming gas can also be decoupled, albeit for a different reason: a significant fraction of molecular gas can be supported by strong turbulence making it less efficient in forming stars \citep[e.g.,][]{semenov19}.

These results signify that the models where star formation is tied to molecular gas should be interpreted with caution. Before disk formation, such modeling is challenging because molecular and star-forming gas phases are decoupled. After disk formation, the two become correlated, however, the total depletion times of molecular gas are significantly shorter than in typical nearby star-forming galaxies. This implies that H$_2$-based star formation models calibrated against local galaxies should not be extrapolated to the early universe; instead, the evolution of the molecular depletion time must be taken into account.

%-----------------------------------------------------------------
%-----------------------------------------------------------------
\section{Summary and Conclusions}
\label{sec:summary}
%-----------------------------------------------------------------

Recent discoveries of a surprisingly high abundance of UV-bright galaxies at early times and the early formation of galactic disks facilitated by JWST, ALMA, and local Galactic archeology surveys pose a challenge to galaxy formation theory. This is especially the case when both are considered simultaneously, as galaxy simulations that can explain high-redshift galaxy luminosity functions via high burstiness of early SFRs also lead to the delayed formation of galactic disks, potentially due to overly vigorous effects of stellar feedback at the early times. 

In the context of these findings, we have investigated the early ($z > 2$) evolution of an early-forming MW-mass disk galaxy in a cosmological zoom-in simulation with detailed modeling of ISM, star formation, and feedback introduced in \citetalias{semenov24a}. The initial conditions were extracted from the TNG50 cosmological-volume simulation, using a close MW analog in terms of the chemo-kinematic properties of the stellar disk at $z=0$ and the timing of disk formation as our target galaxy \citep{semenov23a,chandra23}. Resimulation of this galaxy with explicit modeling of cold turbulent ISM leads to significantly earlier, more efficient, and burstier star formation history and disk formation at a very early time, $z \sim 6\text{--}7$ compared to $z \sim 3$ in the original TNG50 simulation (see \citetalias{semenov24a}). Here we investigate these effects in more detail, in particular, highlighting the importance of modeling locally variable star formation efficiency per freefall time, $\epsff$, coupled with the predicted turbulent state of the ISM on the scales of star-forming regions (see Section~\ref{sec:methods}). Our results are summarized as follows:

\begin{enumerate}
    \item In our simulated galaxy, the long-term evolution of the global SFR is primarily regulated by the evolution of the total mass and average density of the star-forming gas reservoir. The short-term SFR variation is dominated by the variation in the average $\epsff$ as a result of stochastic changes in the turbulent properties of the star-forming gas. This is a new channel for SFR variability that is qualitatively different from the burstiness reported in cosmological simulations with fixed $\epsff$ (Figures~\ref{fig:sfh}--\ref{fig:sfr-scatter}).

    \item The variations of local $\epsff$ exhibit two qualitatively different regimes before and after disk formation. Before disk formation, the $\epsff$ values are systematically higher and its variation is larger in magnitude. This is because violent processes like active gas accretion in cold streams and galaxy mergers drive a small fraction of gas to the star-forming state, but this gas has high $\epsff > 10\%$. After disk formation, the $\epsff$ values and their variation decrease. In this stage, a significant fraction of gas mass resides in the cold state with a fraction of it driven into the star-forming state with low $\epsff < 10\%$ by local turbulence dissipation and disk instabilities (Figure~\ref{fig:nsigma}; see also Figure~\ref{fig:maps} for visual impression).

    \item In both regimes, the $\epsff$ averaged over all star-forming gas stays close to $\epsff \sim 1\%$ which is similar to the value derived for the MW and nearby star-forming galaxies (Figures~\ref{fig:sfh}). This near-universality of average $\epsff$ results from the efficient dispersal of high-$\epsff$ regions by stellar feedback \citep{polzin24}.

    \item By rerunning our galaxy simulation with different fixed $\epsff$ values, we explicitly show that models with low $\epsff$ lead to smaller variations of global SFR and enable the survival of the early disk, while the models with high local $\epsff$ produce more bursty SFRs but also significantly thicken or prevent the formation of the early disk (Figures~\ref{fig:maps-vareff} and \ref{fig:sfh-vareff}).

    \item In our simulation, we explicitly model the formation and destruction of molecular gas. We find that, after the disk formation, the molecular and star-forming gas reservoirs correlate but are decoupled at earlier times, in agreement with previous arguments in the literature \citep[e.g.,][]{clark05,clark18,polzin23}. The depletion times of molecular gas after disk formation are short, 100--300 Myr, despite low local $\epsff \sim 0.5\%$, indicating that the early disk is in the starburst regime driven by high densities of star-forming gas rather than high $\epsff$ (Figure~\ref{fig:Mgas-tdep}).

    \item The depletion time of total ISM gas drops exponentially prior to the disk formation, as the galactic gas reservoir is built up by active accretion and mergers, while after the disk forms, the depletion time stabilizes and gradually increases  (Figure~\ref{fig:Mgas-tdep}). Such a transition from globally inefficient to active star formation is qualitatively similar to the picture suggested by \citet{conroy22} to explain the $\alpha$-element abundances of the low-metallicity MW stars, although the values of depletion times in both stages are $\sim 10$ times shorter in our simulation than these authors report. Note that in our simulation, this increase of global star formation efficiency is driven by the evolution of the global gas reservoir rather than local $\epsff$, as the latter exhibits the opposite trend (see conclusion 2 above).
\end{enumerate}

Our results signify the importance of modeling local variations of the star formation efficiency as opposed to using more common prescriptions with fixed universal values. Local variations of SFE introduce a new physical mechanism for the global SFR variability, which enables the galaxy to transition from highly bursty early star formation to a more steady evolution after the disk formation. This transition enables our simulations to both produce sufficiently large early SFR variations to explain the large abundance of early UV-bright galaxies discovered by JWST, while at the same time enabling the survival of the early forming disk.
The intrinsic trade-off between the early variability of star formation and disk survival implies that considering both simultaneously is a stringent test of star formation and feedback modeling. Therefore in future work, it will be interesting to investigate the performance of the model in a representative sample of zoom-in simulations with different ICs and in large cosmological volumes.

\section*{Acknowledgements}
We are deeply grateful to Oscar Agertz, Renyue Cen, and Andrey Kravtsov for the insightful discussions and comments. 
The analyses presented in this paper were performed on the FASRC Cannon cluster supported by the FAS Division of Science Research Computing Group at Harvard University.
Support for V.S. was provided by Harvard University through the Institute for Theory and Computation Fellowship. L.H. acknowledges support by the Simons Collaboration on ``Learning the Universe.'' 
Analyses presented in this paper were greatly aided by the following free software packages: {\tt yt} \citep{yt}, {\tt NumPy} \citep{numpy_ndarray}, {\tt SciPy} \citep{scipy}, {\tt Matplotlib} \citep{matplotlib}, and \href{https://github.com/}{GitHub}. We have also used the Astrophysics Data Service (\href{http://adsabs.harvard.edu/abstract_service.html}{ADS}) and \href{https://arxiv.org}{arXiv} preprint repository extensively during this project and writing of the paper.

\bibliographystyle{aasjournal}
\bibliography{}

\end{document}